\documentclass[pre,a4paper,twocolumn,showpacs,floatfix]{revtex4-1}
	
    \usepackage[utf8]{inputenc}
    \usepackage[T1]{fontenc}

    \usepackage{amssymb}
    \usepackage{amsmath}
    \usepackage{amsbsy}
    \usepackage{bm}
    \usepackage{graphicx}

\begin{document}
\title{Vicinal surface growth: bunching and meandering instabilities}
\author{Alberto Verga}\email{Alberto.Verga@univ-amu.fr}
   \altaffiliation{%
    IM2NP, CNRS-UMR 7334, France.
   }
   \affiliation{%
   Université d'Aix-Marseille, IM2NP, 
   Campus de St Jérôme, Case 142, 13397 Marseille, France}

\date{\today}

\begin{abstract}
The morphology of a growing crystal surface is studied in the case of an unstable two-dimensional step flow. Competition between bunching and meandering of steps leads to a variety of patterns characterized by their respective instability growth rates. The roughness exponent is shown to go from 1/2 to 1, between the pure bunching to the meandering regimes. Using numerical simulations, we observe that generically, a transition between the two regimes occurs. We find surface shapes and roughness time evolution in quantitative agreement with experiments.
\end{abstract}


\maketitle

The interplay of instability and nonlinearity is central to the formation of spatiotemporal patterns in physical systems driven away from equilibrium \cite{Cross-1993fk}. The typical scenario for the appearance of a periodic structure in a fluid, a spiral in an excitable medium, or a specific front velocity in a reaction-diffusion system, start with the development of a linear instability, which fixes the characteristic scale, followed by some nonlinear saturation mechanism, which fixes the characteristic similarity exponents. This scenario can drastically change in the presence of two simultaneous linear instabilities, because of the competition between different time and length scales.

The step flow \cite{Landau-1950kx,Misbah-2010fk} during molecular beam epitaxy of a vicinal surface is subject to different kinds of instabilities such as bunching (\(x\)-direction) \cite{tersoff95} and meandering (\(y\)-direction) \cite{Bales-1990zl} of steps. The intrinsic anisotropy of the vicinal surface leads to different diffusion and attachment mechanisms and ultimately, to coarsening power laws with different scalings  according to the directions along, or perpendicular to the steps \cite{Verga-2009kx,Keller-2011kx}. Under the simplest conditions of homoepitaxy, bunching and meandering develop together, as for instance in copper \cite{neel03,Yu-2006mz,Yu-2011fk}, and (001) silicon surfaces \cite{schelling99,myslivecek02b,Pascale-2006kx,Frisch-2005ud,Frisch-2006zp}, and are at the origin of a rich variety of surface patterns depending on the growing conditions.

In this Brief Report, we propose a ``vicinal equation'' to describe, in a long wavelength and small amplitude approximation \cite{xiang04}, the simultaneous development of both instabilities \cite{Frisch-2005ud,Frisch-2006zp}. The evolution of the surface height \(z=Ft-mx+h(x,y,t)\), with \(F\) the deposition flux, \(m\) the mean slope and \(h\) the surface shape, is given by,
\begin{equation}
\label{ve}
h_t = \hat L h-\nabla^2|\nabla h|^2-\lambda \nabla\cdot(h_x\nabla h)\,,
\end{equation}
where the operator \(\hat{L}\) is defined by the expression
\[
\hat{L}=(a +\mathrm{i}k_x- k_x^2 -k_y^2)k_x^2 + (b -k_y^2)k_y^2\,,
\]
in Fourier space. The system's behavior is determined by the intrinsic anisotropy of the vicinal surface, and depends on the respective characteristic growth times, fixed by two parameters, \(a\) and \(b\), proportional to the bunching and meandering amplification rates, respectively. The first term, in \(\hat L\), contains the linear effects describing the instabilities, step flow dispersion and diffusion; the second term is the coarsening nonlinearity and the last one the stepping nonlinearity.

\begin{figure}[b]
\includegraphics[width=.16\textwidth]{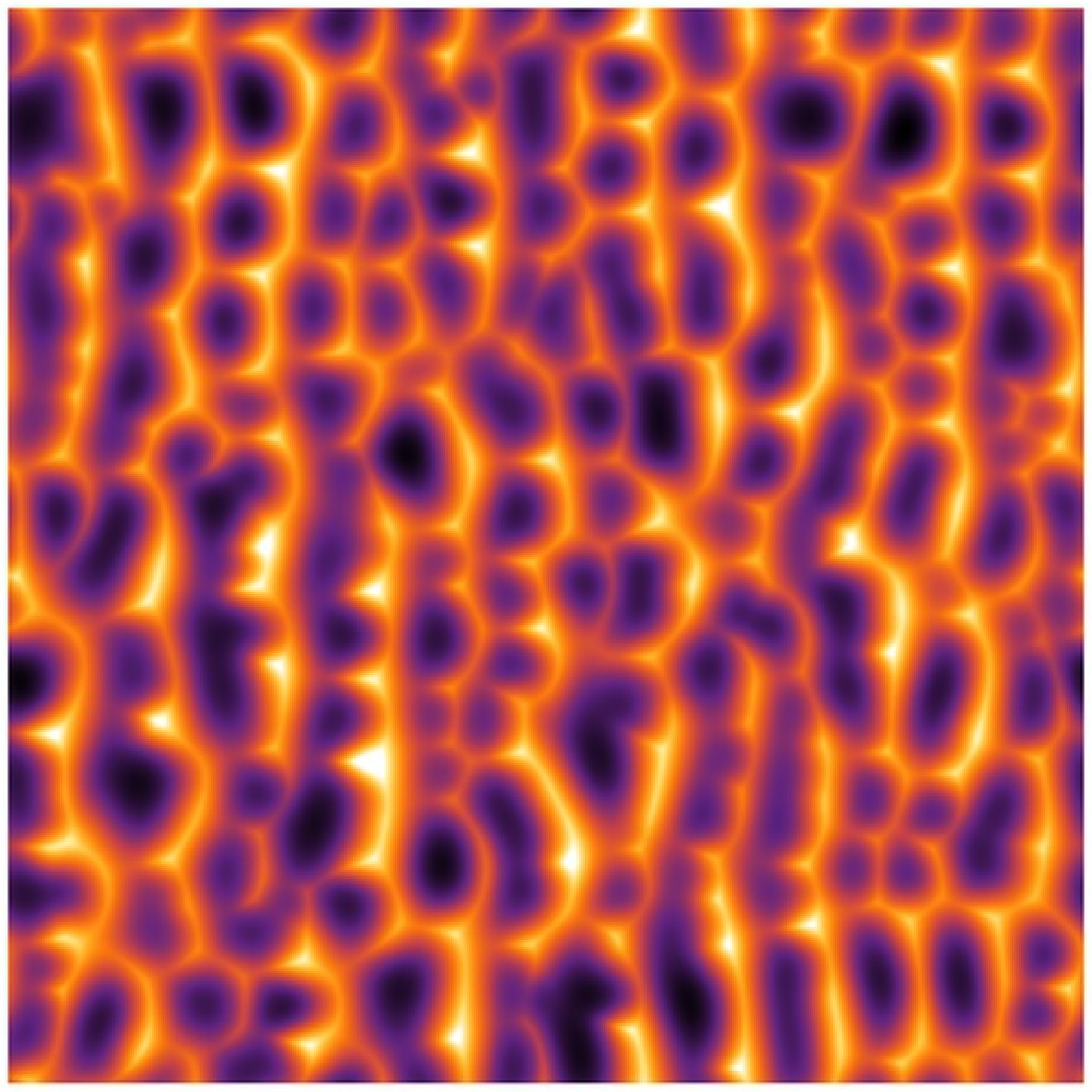}%
\includegraphics[width=.16\textwidth]{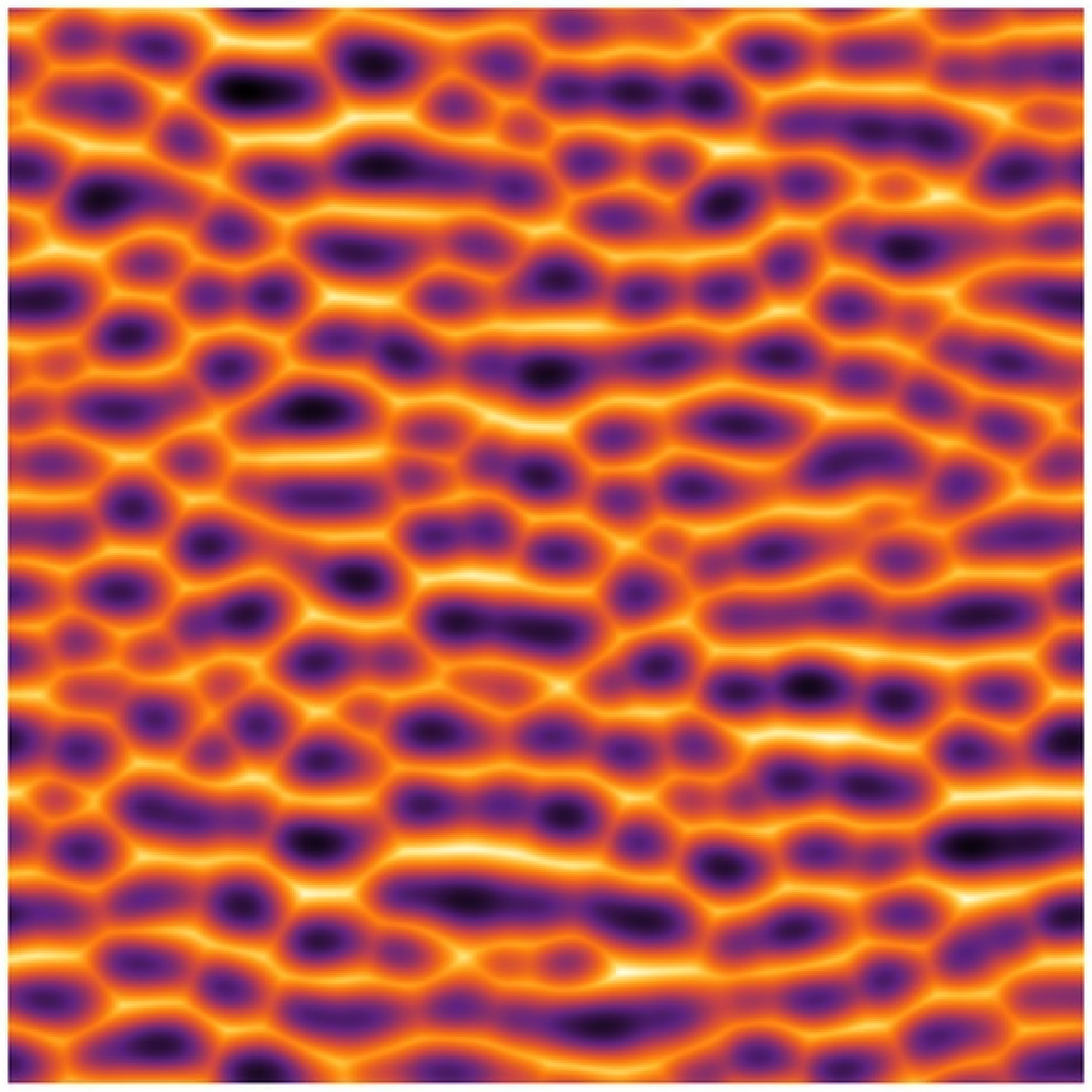}%
\includegraphics[width=.16\textwidth]{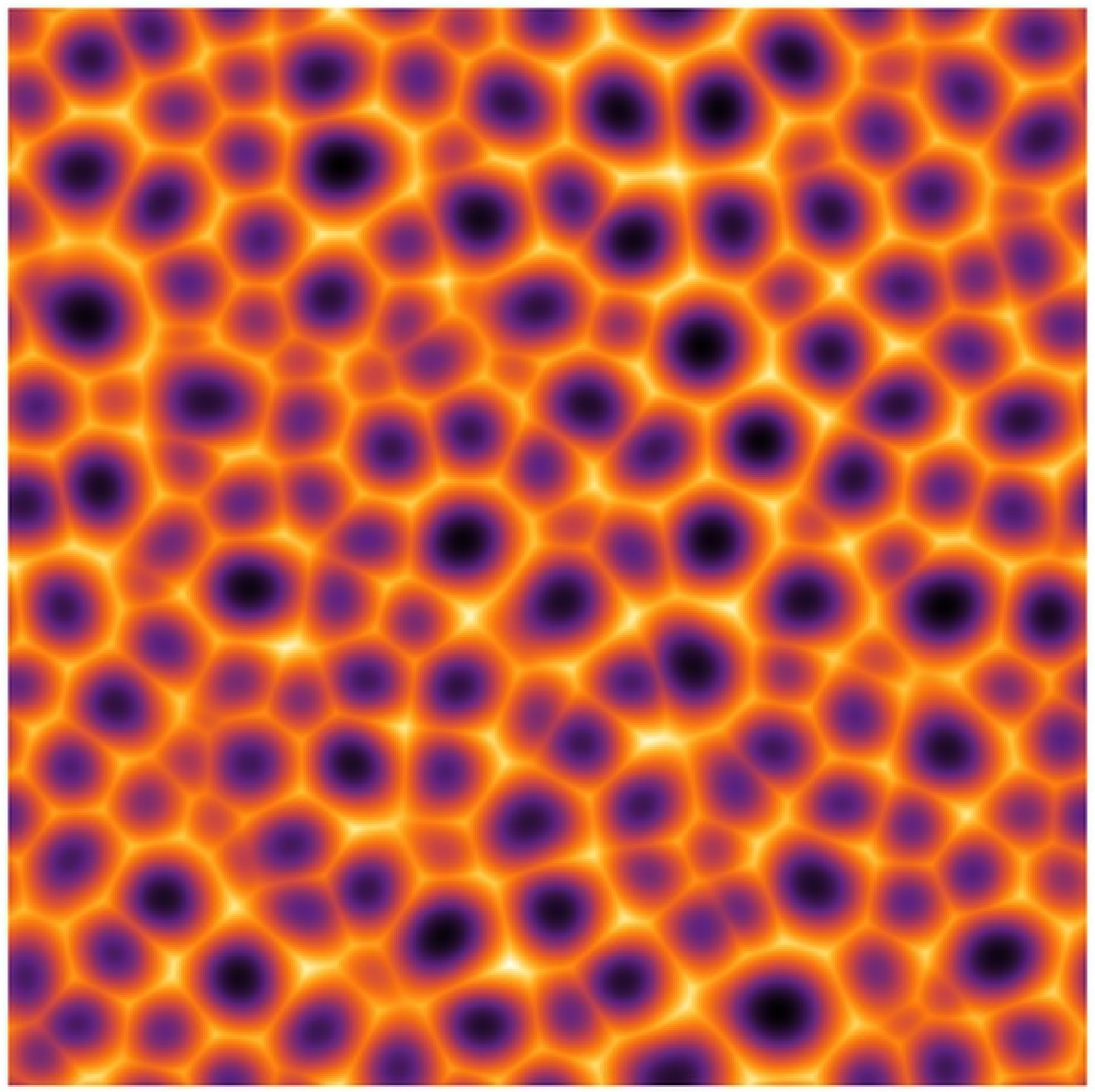}
\caption{\label{BMC} (Color online) Surface patterns viewed from bottom \(-h(\bm{x},t)\); bunches, meanders and cellular patterns, from left to right.}
\end{figure}

\begin{figure*}[t]
\includegraphics[width=.45\textwidth]{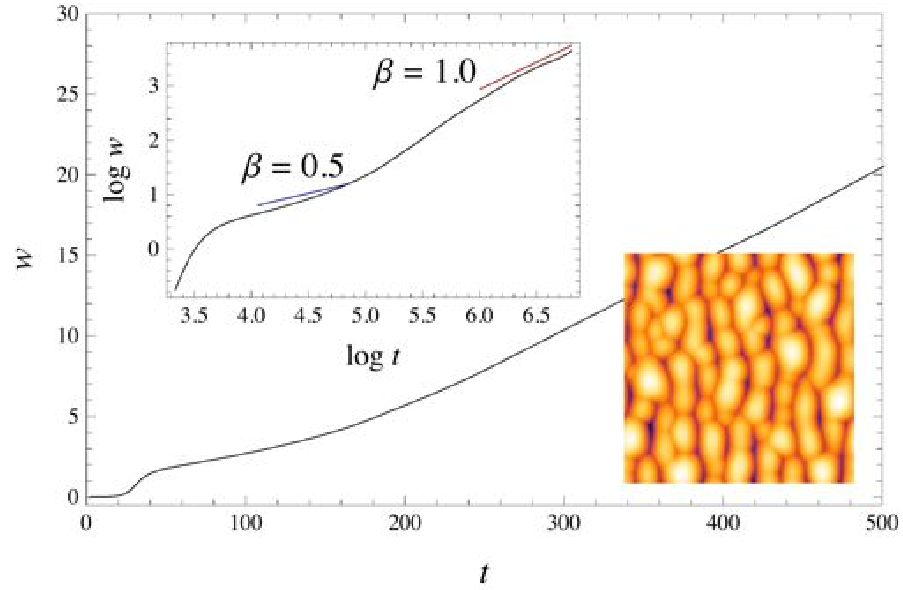}%
\includegraphics[width=.45\textwidth]{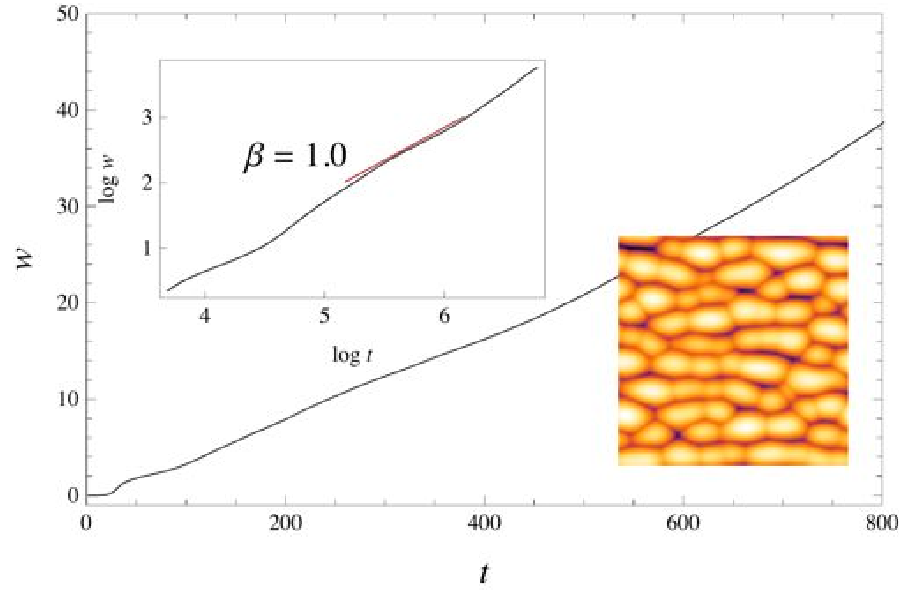}\\
\includegraphics[width=.45\textwidth]{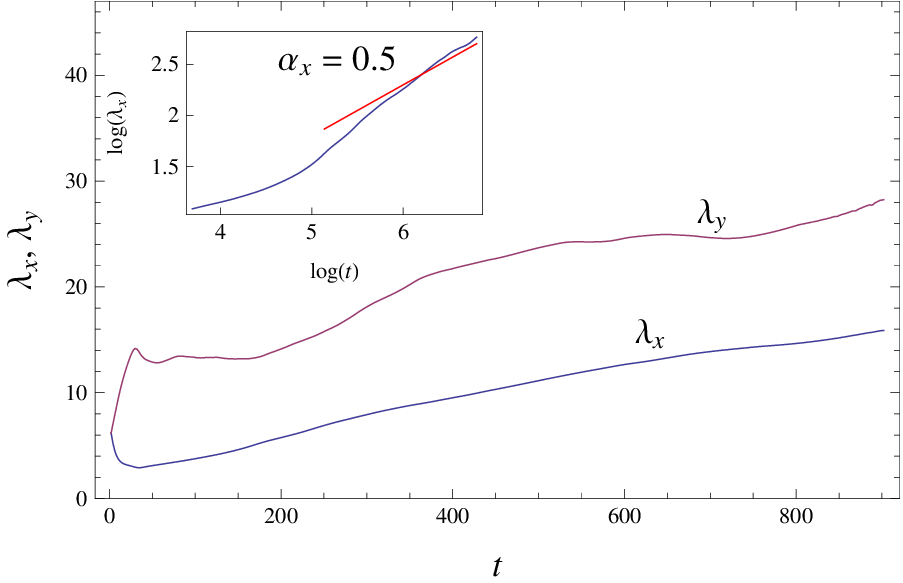}%
\includegraphics[width=.45\textwidth]{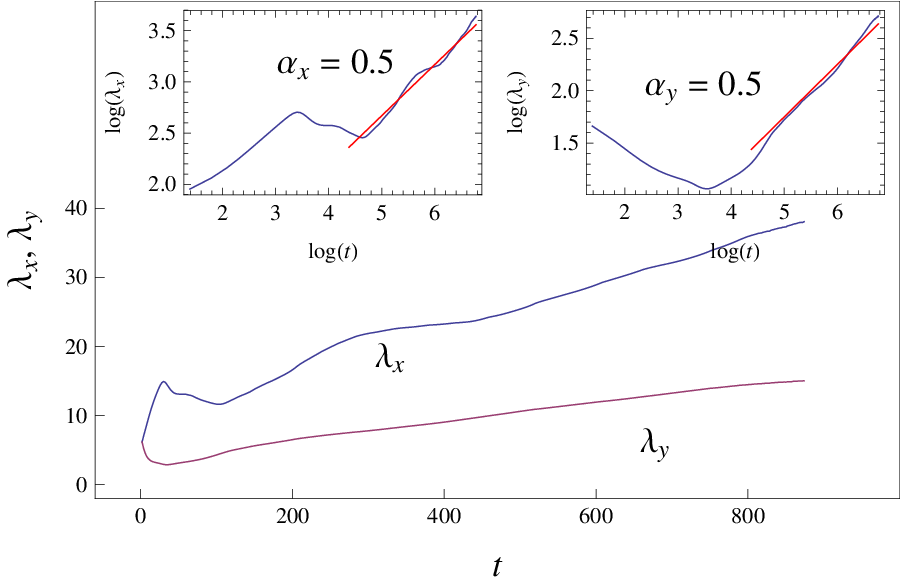}
\caption{\label{f.BMw} (Color online) Bunching instability (left) \(a=1.0\), \(b=0.1\), and meandering instability (right) \(a=0.1\), \(b=1.0\), in the weak stepping nonlinearity regime (\(\lambda=0.2\)). Roughness \(w(t)\) (top) and length scales \(\lambda_x\), \(\lambda_y\) (bottom). Snapshot of the surface shape at \(t=400\) (inset). The similarity exponents \(\beta\) and \(\alpha\) are shown to schematically indicate the different regimes (insets in log-log scale).}
\end{figure*}

Equation (\ref{ve}) amounts to a generalization of the anisotropic evolution of a vicinal surface, unstable with respect to meandering but stable to bunching, of Ref.~\cite{Verga-2009kx}. The essential difference is the presence of a new nonlinearity, the last term in (\ref{ve}), proportional to the coupling parameter \(\lambda\), that breaks the symmetry \(x\leftrightarrow -x\), and allows us to take into account the mechanisms leading to step bunching. It can be viewed as a simple small gradient expansion of step flow driven by a non-equilibrium current \cite{Krug-2002cr},
\begin{equation}
z_t=-\frac{F\delta}{2}\nabla\cdot\left[\frac{\nabla z}{|\nabla z|^2}\right]+F
\end{equation}
where \(\delta=l_{\!+}-l_{\!-}\) is the difference between the up step \(l_{\!+}\) and down step \(l_{\!-}\) attachment lengths; in this case one obtains \(\lambda=F\delta/m^3\), and bunches form if \(\delta<0\). 

We may consider Eq.~(\ref{ve}) as a phenomenological model of the vicinal surface growth; the actual form of the three independent parameters \((a,b,\lambda)\) depends on the physical details of the meandering and bunching instabilities. Its special form assumes that the essential mechanisms of these instabilities are kinetic, as appropriated for homoepitaxy conditions. Throughout this paper we will use units based on the equilibrium terrace length \(l_0\), and time \(l_0^2/D_0C_0\), with adatom concentration \(C_0\) and diffusion coefficient \(D_0\), conveniently rescaled to throw out non relevant dependent parameters \cite{Verga-2009kx}. Therefore, comparison with experiments can be done by choosing length and time scales to fit a given reference state.

For different sets of parameters \(a\), \(b\) and \(\lambda\), meanders, step bunches or cellular patterns may be obtained, as illustrated in Fig.~\ref{BMC}. The snapshots show the surface shape, obtained after the evolution of an initially homogeneous state perturbed by small amplitude noise, viewed from the bottom, to better visualize the strong gradients formed in  both directions.

\begin{figure*}
\includegraphics[width=.45\textwidth]{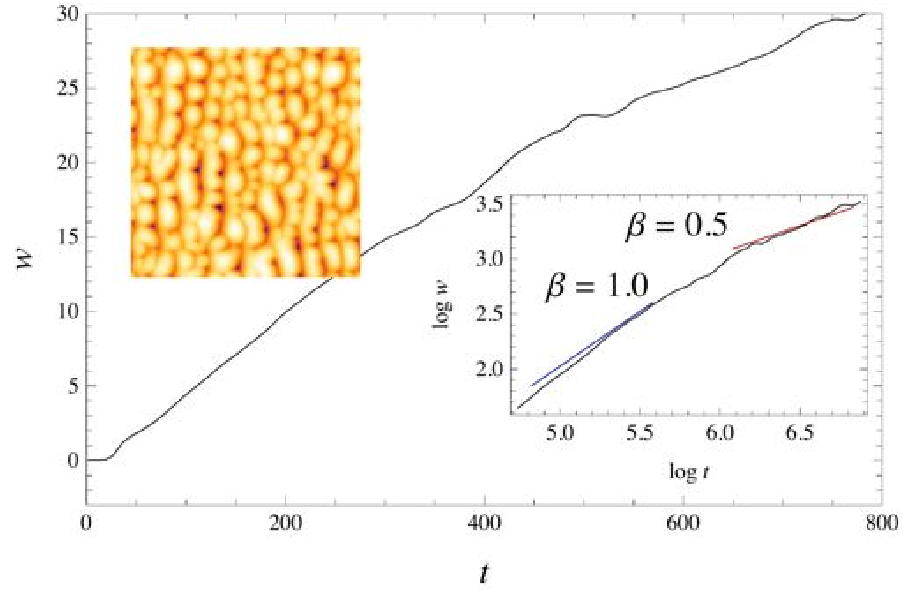}%
\includegraphics[width=.45\textwidth]{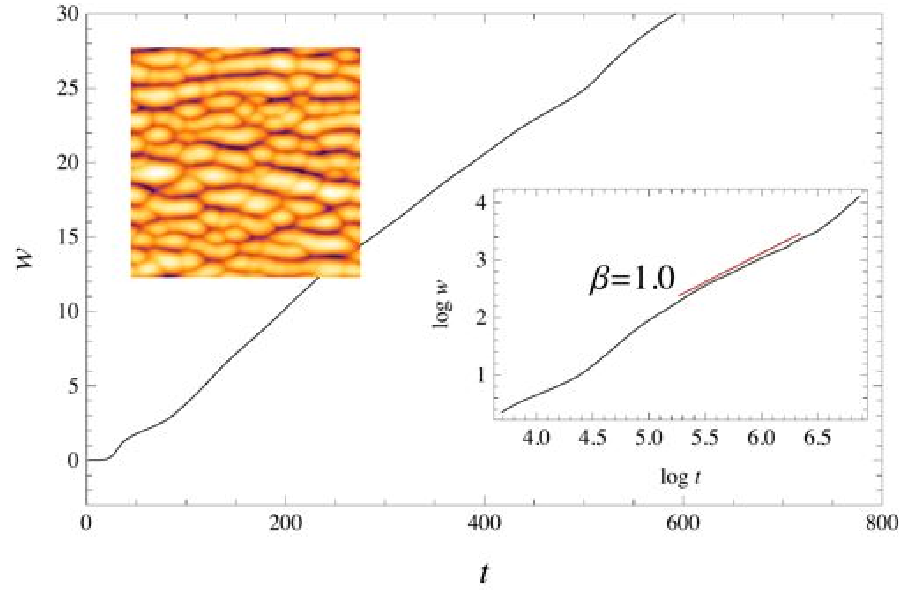}\\
\includegraphics[width=.45\textwidth]{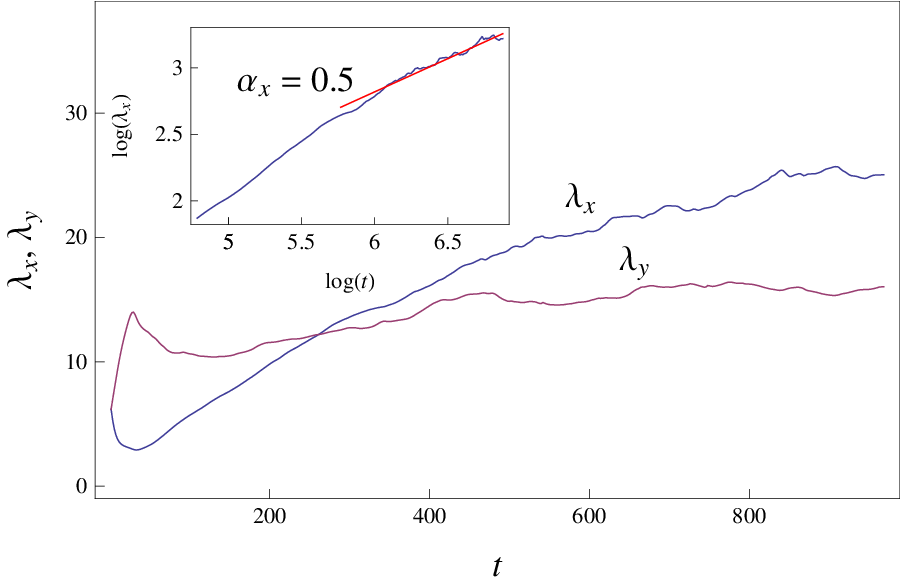}%
\includegraphics[width=.45\textwidth]{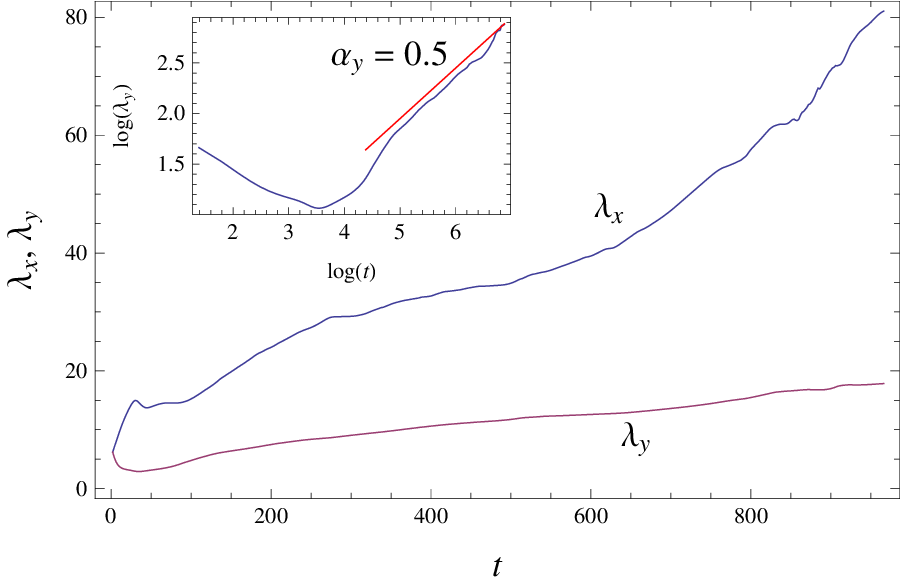}
\caption{\label{f.BMs} (Color online) Bunching instability (left) \(a=1.0\), \(b=0.1\), and meandering instability (right) \(a=0.1\), \(b=1.0\), in the strong stepping nonlinearity regime (\(\lambda=1.0\)). Roughness \(w(t)\) (top) and length scales \(\lambda_x\), \(\lambda_y\) (bottom). Snapshot of the surface shape at \(t=160\) (inset). The similarity exponents \(\beta\) and \(\alpha\) are shown to schematically indicate the different regimes (insets in log-log scale).}
\end{figure*}

We can distinguish two limiting one-dimensional approximations according to the relevant unstable direction. In the bunching dominated regime the effective equation is,
\begin{equation}\label{xe}
h_t\sim -a h_{xx}-\lambda (h_x^2)_x
\end{equation}
(we use subscripts to denote partial derivatives). Putting
\begin{equation}\label{e.eB}
h(x,t)=t^\beta H(x/t^{\alpha_x})\,,
\end{equation}
one obtains \(\alpha_x=\beta=1/2\), where the exponents \(\alpha_x\) and \(\beta\) govern the characteristic bunching length and the roughness time scales, respectively. In the meandering regime we have,
\begin{equation}\label{ye}
h_t\sim -b h_{yy}-(h_y^2)_{yy}
\end{equation}
and 
\begin{equation}\label{e.eM}
h(y,t)=t^\beta H(y/t^{\alpha_y})
\end{equation}
which gives \(\beta=1\) for the roughness exponent and \(\alpha_y=1/2\), for the meander length scale. In addition, the stable direction is governed by the bi-quadratic equation \(h_t\sim-\nabla^4 h\), leading to a characteristic spreading exponent \(1/4\), in the \(x\)-direction for bunching, and in the \(y\)-direction for meandering. Therefore, one may predict that the roughness exponent \(\beta\) should go from \(1/2\) in the pure bunching, to \(1\) in the meandering regime. We performed a series of numerical computations to obtain the asymptotic behavior of the surface roughness, and to investigate the pattern formation in the presence of the anisotropic instabilities. Our goal is to compare these results with experiments of Si (001) homoepitaxy \cite{schelling99,myslivecek02b,Pascale-2006kx}.

To illustrate the two limiting regimes, we choose two sets of linear parameters: (i) \(a=1.0\), \(b=0.1\) for the bunching case, and (ii) \(a=0.1\), \(b=1.0\) for the meandering case. The actual behavior of the system also depends on \(\lambda\), we distinguish between a weak nonlinearity case \(\lambda=0.2\), and a strong one \(\lambda=1.0\). The system size is a \(256\times 256\) square (we use a pseudo-spectral algorithm with \(512^2\) collocation points). Results in these two regimes are displayed in Figs.~\ref{f.BMw} and \ref{f.BMs}. We plot the roughness,
\begin{equation}
w(t)=\sqrt{\langle h(t)^2\rangle -\langle h(t)\rangle^2}\,,
\end{equation}
where the average is over the whole surface, and the characteristic lengths,
\begin{equation}
(\lambda_x^{-2},\lambda_y^{-2})=
\frac{\int d\bm k \,(k_x^2,k_y^2)\,\left|h_{\bm k}\right|^2}{\int d\bm k \left|h_{\bm k}\right|^2}\,,
\end{equation}
where \(h_{\bm k}=h_{\bm k}(t)\) is the two-dimensional Fourier transform, together with the corresponding snapshots of the surface shape. The lengths \(\lambda_y(t)\) and \(\lambda_x(t)\) are the characteristic scales in the meander and the step gradient directions, respectively.

\begin{figure*}
\includegraphics[width=.23\textwidth]{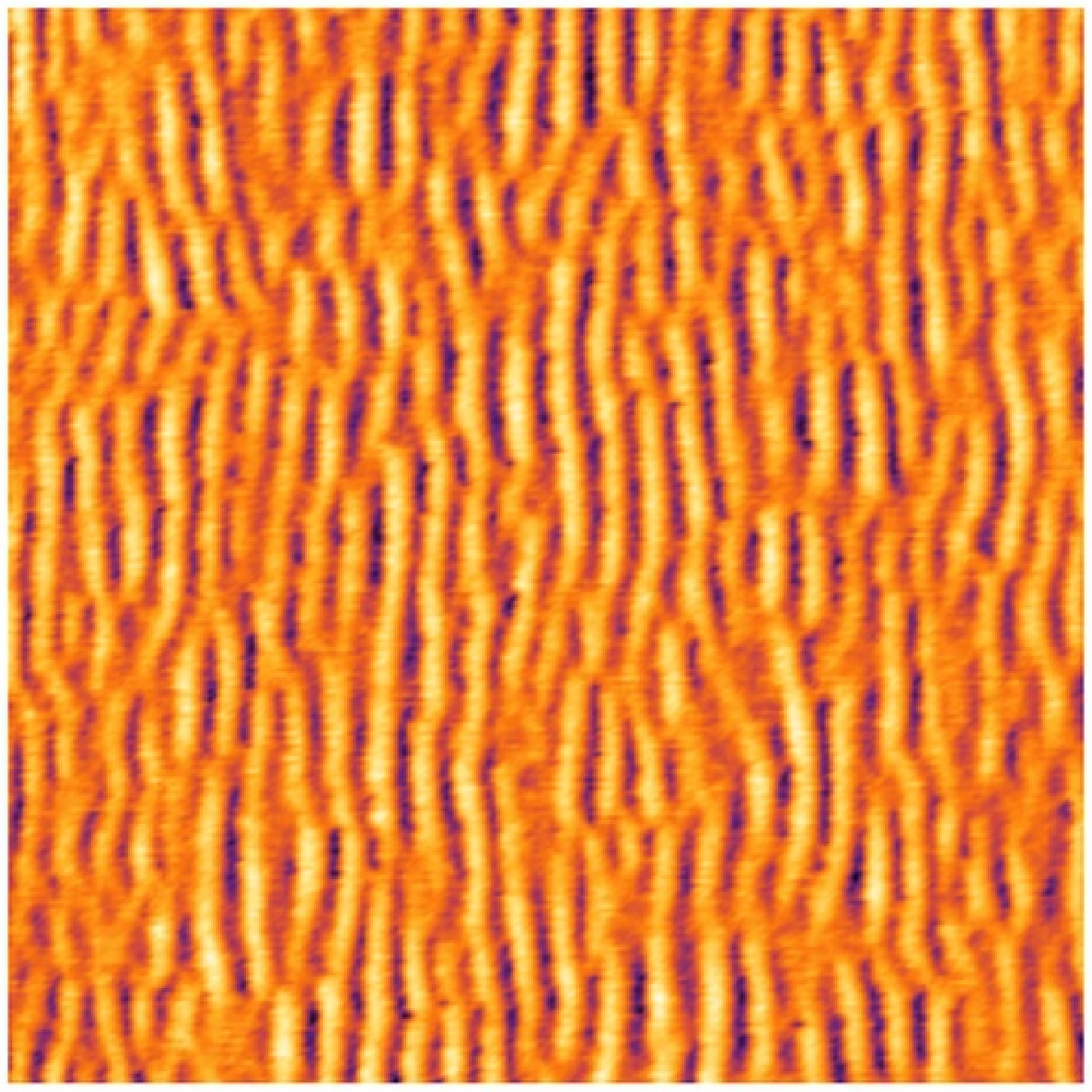}%
\includegraphics[width=.23\textwidth]{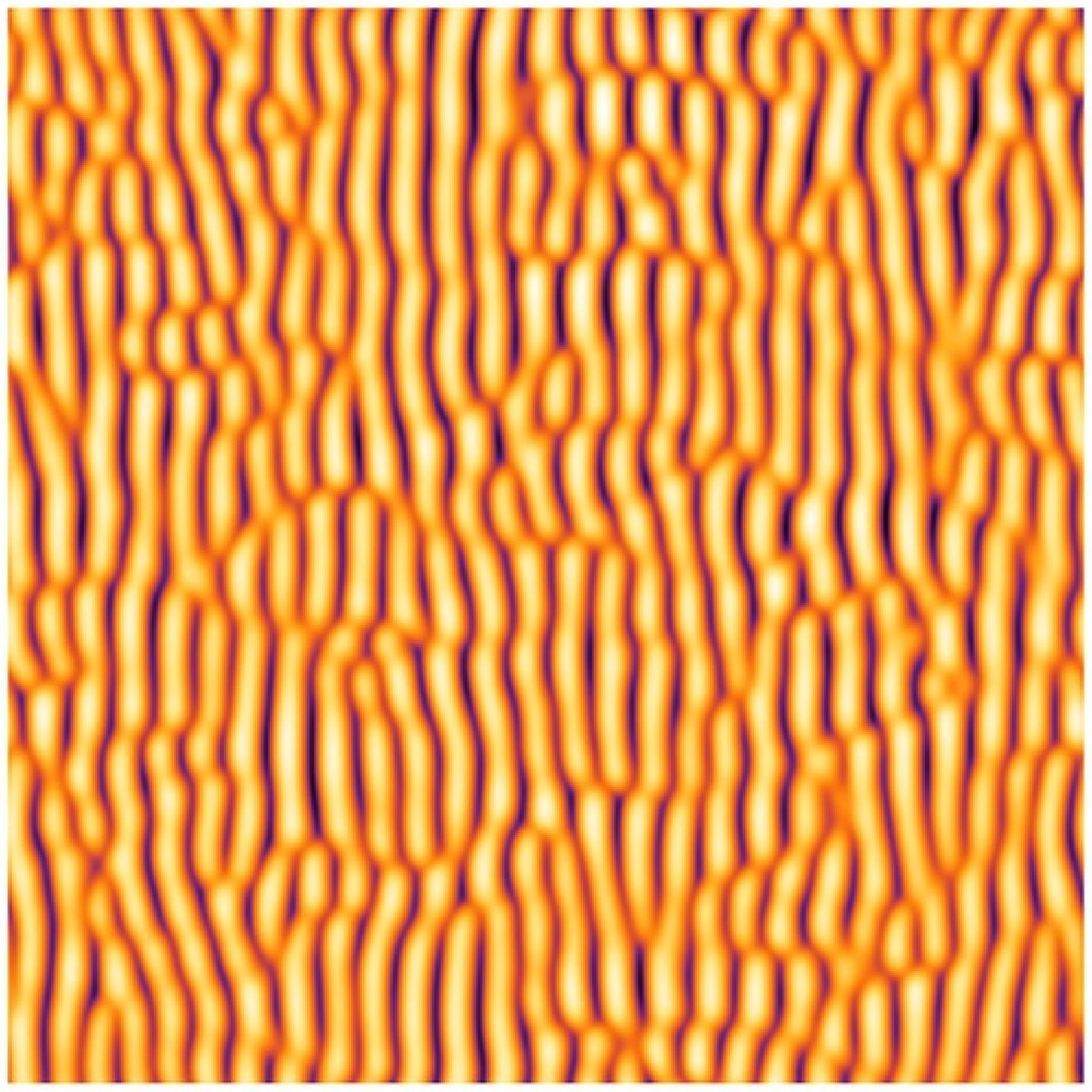}%
\includegraphics[width=.23\textwidth]{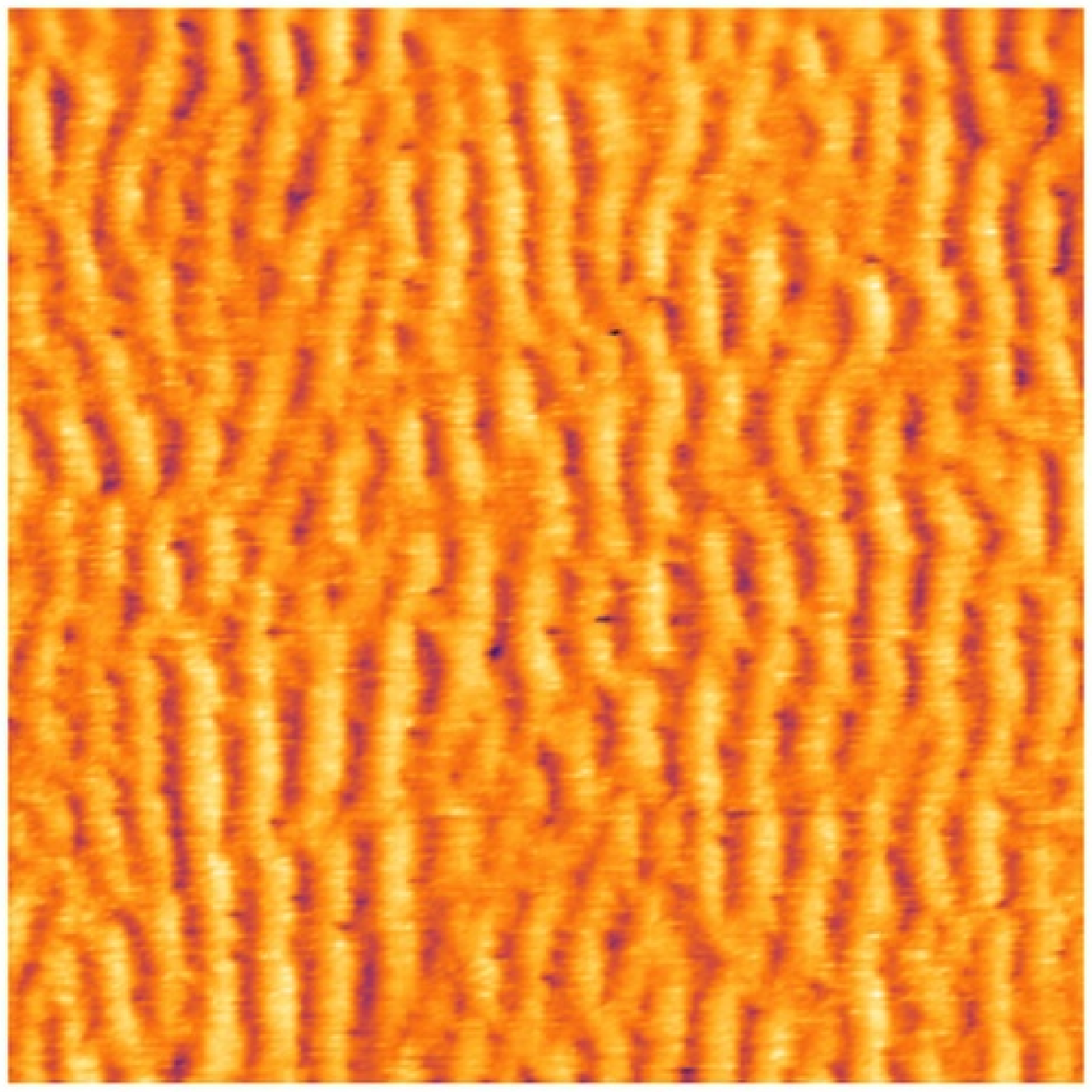}%
\includegraphics[width=.23\textwidth]{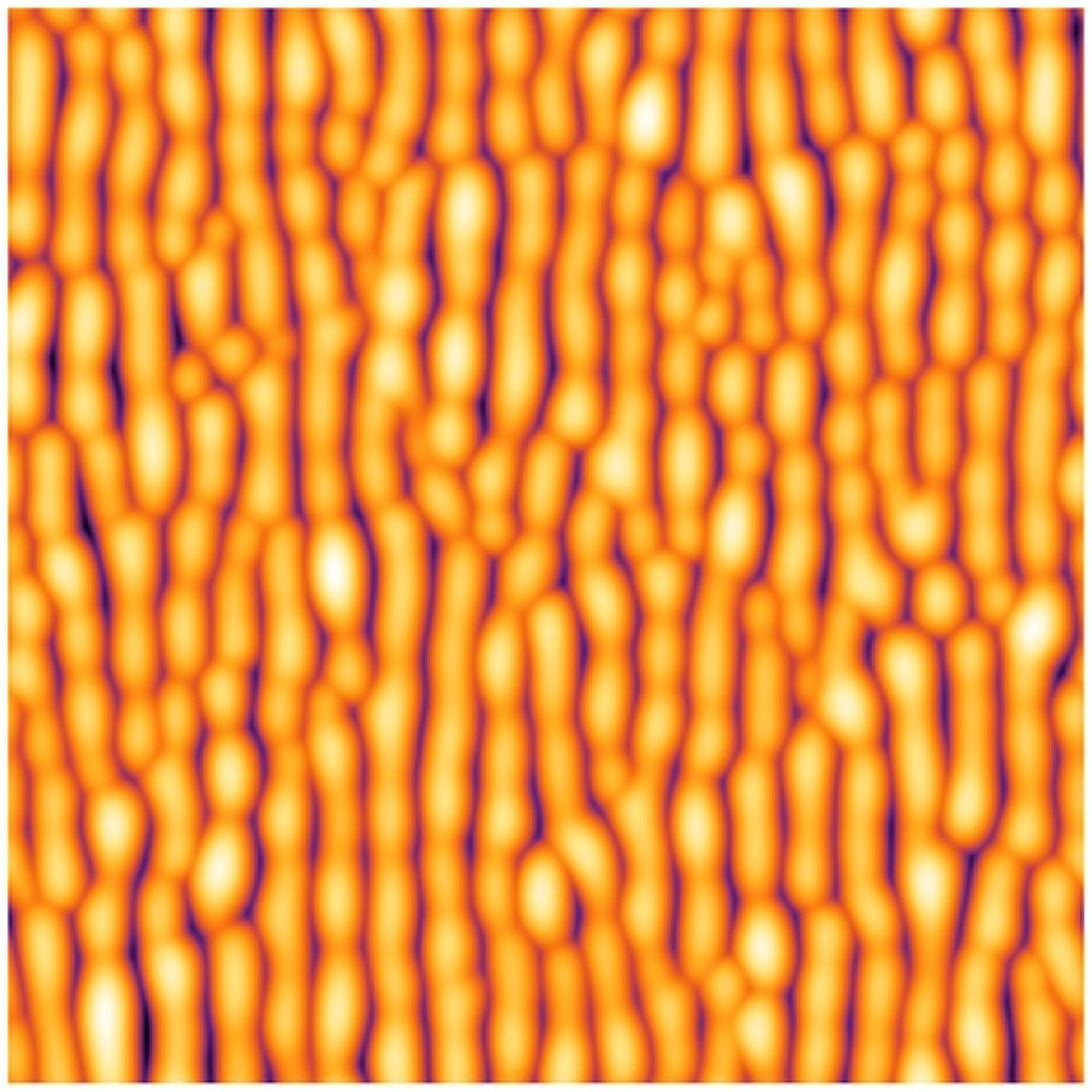}
\caption{\label{EN} (Color online) Comparison of experimental AFM images and numerical simulations. From left to right, AFM at thickness \(100\), model at \(t=20\), AFM at thickness \(300\), and model at \(t=60\). Numerical parameters as in Fig.~\protect\ref{f.BMw} (left).}
\end{figure*}

In accordance with (\ref{e.eB}) and (\ref{e.eM}), we may expect the asymptotic behaviors: \(w\sim t^\beta\) and \(\lambda_x\sim t^{\alpha_x}\), \(\lambda_y\sim t^{\alpha_y}\). One may think in particular that, in the bunching case, the roughness evolution actually observed would start, after a transitory dependence on the initial state, with a \(w\approx t^{1/2}\) growth (dominant at small times), followed by a \(w\approx t\) at larger times. This is effectively the case when the stepping nonlinearity is dominated by coarsening (Fig.~\ref{f.BMw}, top-left). Indeed, after an initial evolution characterized by the bunching instability with \(\beta=1/2\), the coarsening regime is established at long-times, leading to a roughness evolution with \(\beta=1\).

However, at long times, depending on the value of \(\lambda\), a new regime may arise. It depends on the relative strength of the coarsening nonlinearity and the stepping nonlinearity. These become comparable at the scale:
\[
\nabla^2|\nabla h|^2 \sim \lambda\nabla\cdot(h_x\,\nabla h)\,.
\]
Assuming \(\lambda_y\sim \lambda_x\sim l \ll L\) to ensure that two-dimensional effects are important, this can be satisfied if
\[
\frac{1}{l^2} \sim \frac{\lambda}{l}
\]
where \(L\) is the system size. In the step bunching unstable flow, when \(\lambda l(t)\) becomes greater than one, the stepping nonlinearity overcomes the coarsening one. In this case the roughness would be characterized by an exponent \(\beta=1/2\), as found in the strong bunching regime (Fig.~\ref{f.BMs}, top-left). We then observe an initial regime dominated by coarsening (with \(\beta=1\)), followed by a regime dominated by the stepping nonlinearity (with \(\beta=1/2\)). It is worth noting that the change in regime takes place around \(t\approx 300\), when the \(\lambda_x\) scale overcomes the \(\lambda_y\) scale; in the weak stepping nonlinearity regime of Fig.~\ref{f.BMw}, this never happens.

In the case where \(b>a\) the meandering instability dominates (Figs.~\ref{f.BMw} and \ref{f.BMs}, right). For both cases, weak and strong stepping nonlinearity, the \(\lambda_x(t)\) characteristic length rapidly increases, signaling that a quasi one-dimensional state sets in. The roughness follows a power law with \(\beta=1\). The phenomenology of the meander coarse graining is similar to the one described by the weak nonlinear amplitude equation for the Bales-Zangwill instability \cite{Frisch-2007zl,Verga-2009kx}. It is interesting to note the meander morphology is modified by the presence of a bunching instability, although weak, that accelerates the homogenization along the steps, eliminating the gradients in this direction.

In the bunching regime \(a>b\), for weak (but unstable) meandering and step nonlinearity, the shape of the surface consists on a succession of bunches with some dislocation type defects. It compares satisfactorily with the atomic force microscopy (AFM) measurements of a silicon vicinal \((001)\) surface \cite{Pascale-2006kx}, as shown in Fig.~\ref{EN}. The observed quasi-periodic pattern of step bunches and the time scales are well reproduced by the numerical model (\ref{ve}). In particular, the observed experimental evolution of the roughness, that the model predicts to start with a first asymptotic with \(\beta=1/2\) afterward increasing up to \(\beta=1\), actually shows a breaking between a slow initial grow that later becomes steeper; a fit over the full time, mixing the two regimes, gives an intermediate exponent \(\beta_{exp}=0.53\).

In this Brief Report we presented a weak nonlinear model of the evolution of a vicinal surface subject simultaneously to bunching and meandering of steps. We found that depending on the relevant nonlinear mechanism, coarsening or stepping, different long time evolutions of the surface roughness are possible. We also showed that the presence of a weak, subdominant, bunching instability, modifies the morphology of the meanders, leading to a quasi one-dimensional pattern; the same is true for the bunching dominant regime, when coarsening is the relevant nonlinear effect. In conclusion, the phenomenological vicinal surface equation is able to reproduce the observed morphology and roughness evolution of homoepitaxial growth, and provides an useful framework for investigating patterns in anisotropic unstable systems.

I thank T. Frisch, R. Cuerno and P. Tejedor for valuable discussions; I also acknowledge I. Berbezier and A. Ronda, who provided me the data used in Fig.~\ref{EN}.


%

\end{document}